\documentclass[twocolumn,showpacs,amssymb,aps,prc,nofootinbib,floatfix,superscriptaddress]{revtex4-1}
\usepackage{amsmath,graphicx,color,ulem}
\usepackage{hyperref}
\newcommand{\trento}{T$\mathrel{\protect\raisebox{-2.1pt}{R}}$ENTo}

\begin{document}
\title{Extracting the speed of sound of QCD from transverse momentum fluctuations}

\author{Mubarak Alqahtani}
\affiliation{Department of Physics, College of Science, Imam Abdulrahman Bin Faisal University, Dammam 31441, Saudi Arabia  }
\affiliation{Basic and Applied Scientific Research Center, Imam Abdulrahman Bin Faisal University, Dammam 31441 Saudi Arabia}
\author{Tribhuban Parida}
\affiliation{AGH University of Krakow, Faculty of Physics and Applied Computer Science, aleja Mickiewicza 30, 30-059 Cracow, Poland}
\author{Jean-Yves Ollitrault}
\affiliation{Universit\'e Paris Saclay, CNRS, CEA, Institut de physique th\'eorique, 91191 Gif-sur-Yvette, France}

\begin{abstract} 
We extract the speed of sound ($c_s$) in the quark-gluon plasma from ATLAS data on the probability distribution of the transverse momentum per particle, $[p_T]$, in ultra-central Pb+Pb collisions.  
With an ideal detector, $c_s$ can be inferred from the rise of the mean $[p_T]$ with the collision multiplicity. 
In practice, however, low-$p_T$ particles escape detection, which biases the analysis. 
We show how to correct for this bias by using data on the variance of $[p_T]$, as well as information from the recently-measured $v_0(p_T)$. 
We also introduce a systematic method for deblurring the noise from the hadronization process. 
Assuming that the size of the quark-gluon plasma is independent of the hadron multiplicity in  collisions at zero impact parameter, which is the scenario preferred both by high-energy QCD and heavy-ion data, we obtain  $c_s/c=0.496\pm 0.008$ at temperature $T=221\pm 13$~MeV, in perfect agreement with first-principles calculations from lattice QCD. 

\end{abstract}
\maketitle

\section{Introduction}
\label{s:introduction}

Collisions between heavy nuclei at ultrarelativistic energies create a thermalized fluid, which is interpreted as a droplet of quark-gluon plasma within the theory of strong interactions~\cite{Busza:2018rrf}. 
Thermal equilibrium is only a crude approximation, and the temperature of the fluid is neither uniform nor constant.
However, for a smoothly-varying equation of state, state-of-the-art hydrodynamic simulations have revealed a robust proportionality between the transverse momentum per particle $[p_T]$ and an effective temperature $T_{\rm eff}$~\cite{Gardim:2019xjs,Gardim:2024zvi,SoaresRocha:2024drz,Gavassino:2025bts}, whose definition is recalled in  Sec.~\ref{s:hydropicture}. 
It was predicted on this basis~\cite{Gardim:2019brr} that the mean value of $[p_T]$ would increase as a function of  the number of charged particles $N_{ch}$ in ``ultra-central'' collisions, representing the $\approx 0.4\%$ fraction of collisions with the largest $N_{ch}$, whose impact parameter $b$ is close to $0$.
The physical picture is that more particles in a constant volume imply a higher density, which in turn implies a higher temperature.
This prediction was confirmed in Pb+Pb collisions at the Large Hadron Collider (LHC)~\cite{CMS:2024sgx,ALICE:2025rtg,ATLAS:2024jvf}.
One also observes that the variance of $[p_T]$ across collision events with the same $N_{ch}$ {\it decreases\/} in ultra-central collisions. 
This phenomenon, which had not been predicted, is another consequence of thermalization~\cite{Samanta:2023amp}. 
Event-by-event fluctuations of $[p_T]$ at fixed $N_{ch}$ result from fluctuations in the volume of the quark-gluon plasma, which induce density fluctuations. 
These volume fluctuations largely result from fluctuations of $b$ relative to $N_{ch}$, since $b$ determines the overlap area between the nuclei. 
In ultra-central collisions, however, $b$ becomes closer and closer to $0$, leaving little room for fluctuations, and the variance of $[p_T]$ decreases. 

In this paper, we extract a quantitative estimate of the speed of sound $c_s$ at temperature $T_{\rm eff}$  from recent ATLAS data~\cite{ATLAS:2024jvf}.
$c_s$ is defined from the increase of temperature with density.
It has originally been proposed to infer its value from the increase of the mean $[p_T]$ in ultra-central collisions~\cite{Gardim:2019brr,CMS:2024sgx}, as recalled in Sec.~\ref{s:hydropicture}. 
This method has been shown to suffer from a number of biases~\cite{Nijs:2023bzv} which we systematically correct.
ATLAS does not detect particles with $p_T<0.5$~GeV$/c$. 
In Sec.~\ref{s:v0pt}, we take into account this effect, which involves the new observable $v_0(p_T)$~\cite{Schenke:2020uqq,Parida:2024ckk,ATLAS:2025ztg,ALICE:2025iud}.
Due to this effect, the determination of $c_s$ requires not only the mean, but also the variance of $[p_T]$ (Sec.~\ref{s:gaussian}). 
Fluctuations of impact parameter are evaluated in Sec.~\ref{s:geometry}. 
In Sec.~\ref{s:poisson}, we introduce a systematic procedure to unfold the statistical fluctuations originating from the hadronization process. 
In Sec.~\ref{s:fits}, we present the ATLAS results~\cite{ATLAS:2024jvf} and our fits, from which we infer the value of $c_s$. 
We compare this value with first-principles calculations in Sec.~\ref{s:lattice}.

\section{Hydrodynamic response to initial-state fluctuations}
\label{s:hydropicture}

The two observables of interest for this paper are the particle multiplicity $N$ and the transverse momentum per particle $[p_T]$, which fluctuate event to event.
%%In a hydrodynamic simulation, both are evaluated by integrating the momentum distribution over the freeze-out hypersurface. 
We first assume that all the particles emitted around mid-rapidity are detected, irrespective of their transverse momentum $p_T$. 
This assumption will be relaxed in Sec.~\ref{s:v0pt}. 
We denote by $N$ the mean multiplicity of a hydrodynamic event, which can vary continuously. 
Discretization effects will be thoroughly modeled in Secs.~\ref{s:geometry} and \ref{s:poisson}, and ignored until then. 

Since the equations of hydrodynamics are deterministic, $N$ and $[p_T]$ solely depend on the initial conditions of the hydrodynamic calculation, typically at an early time $\tau_0$. 
We introduce a simplified effective description which gives a good approximation to full numerical results~\cite{Gardim:2019xjs,Gardim:2020sma}. 
We denote by $s({\bf x})$ the initial boost-invariant~\cite{Bjorken:1982qr} entropy density profile at time $\tau_0$, where ${\bf x}$ is the transverse coordinate. 
The two quantities of interest for the effective description are the entropy per unit rapidity $S$ and the rms transverse radius $R$~\cite{Zhou:2025bwu}: 
\begin{align}
  \label{defS}
  S&\equiv \tau_0\int_{\bf x} s({\bf x})\nonumber\\
  R^2&\equiv \frac{\tau_0}{S} \int_{\bf x}|{\bf x}|^2 s({\bf x}) -\left|\frac{\tau_0}{S} \int_{\bf x}  {\bf x}\, s({\bf x})   \right|^2. 
\end{align}
This initial condition is then evolved through the equations of relativistic viscous hydrodynamics~\cite{Romatschke:2007mq,Gale:2013da}. 
The fluid expands into the vacuum and cools down until it is converted into hadrons. 
The points where this conversion occurs, which come closer to the centre as time evolves~\cite{Luzum:2008cw}, define a freeze-out hypersurface. 
Our effective description relies on the total energy $E_f$ and entropy $S_f$ per unit space-time rapidity, obtained upon integration over this freeze-out hypersurface~\cite{Gardim:2019xjs}. 
The effective temperature $T_{\rm eff}$ and effective volume $V_{\rm eff}$ are defined as those of a fluid at rest with energy $E_f$ and entropy $S_f$: 
\begin{align}
\label{effectivehydro}
E_f&=\epsilon(T_{\rm eff}) V_{\rm eff},\nonumber\\
S_f&=s(T_{\rm eff}) V_{\rm eff},
\end{align}
where $\epsilon$ denotes the energy density, which is related to $s$ through the equation of state used in the hydrodynamic simulation. 
Let us briefly discuss the physical interpretation of $T_{\rm eff}$: 
If the initial density profile $s({\bf x})$ was uniform and if there was no expansion, $T_{\rm eff}$ would simply be the initial temperature. 
Expansion work decreases the energy~\cite{Bjorken:1982qr}, so that  $T_{\rm eff}$ is always smaller than the initial temperature. 
On the other hand, $T_{\rm eff}$ is larger than the temperature at freeze-out, because the final energy $E_f$ includes the kinetic energy of the fluid. 
It can be viewed as the freeze-out temperature, blue shifted by the transverse expansion. 

The final-state observables $N$ and $[p_T]$ are obtained by integrating the phase-space distributions of outgoing hadrons over the freeze-out hypersurface. 
We now sketch how they relate to the initial-state quantities $S$ and $R^2$. 
$N$ is proportional to $S_f$, where the proportionality factor is determined by the hadron resonance gas model~\cite{Hanus:2019fnc}, and $S_f$ is proportional to $S$ ($S_f=S$ in the limit of vanishing viscosity, $S_f>S$ in general).  
As for $[p_T]$, one finds numerically that it is proportional to $T_{\rm eff}$~\cite{Gardim:2024zvi}, which is related to $s(T_{\rm eff})$ through the equation of state. 
Finally, $V_{\rm eff}$ is proportional to $R^3$ for dimensional reasons. 
Putting together these results, we obtain: 
\begin{align}
\label{effhydro}
N&\propto S\nonumber\\
[p_T] &\propto T_{\rm eff}\nonumber\\
s(T_{\rm eff})&\propto \frac{S}{(R^2)^{3/2}},
\end{align}
where the proportionality factors are identical for all events. 
That in the second line will be discussed in detail in Sec.~\ref{s:lattice}, the other two will not be needed in this paper. 

We now study more specifically how event-by-event fluctuations are carried over from the initial to the final state. 
There are two distinct  sources of event-by-event fluctuations: 
Fluctuations of impact parameter $b$, which are classical~\cite{Samanta:2023amp,Roubertie:2025qps}, and fluctuations originating from the wavefunctions of incoming nuclei, which are quantum. 
This distinction is essential in order to determine the speed of sound from $[p_T]$ data~\cite{Gardim:2019brr,Samanta:2023amp}. 
We first study fluctuations at fixed $b$. The average over $b$~\cite{Das:2017ned} will be studied in Sec.~\ref{s:geometry}. 

Thus, we consider an ensemble of events with the same $b$. 
Hydrodynamic simulations show that in this ensemble, Eqs.~(\ref{effhydro}) hold, to a good approximation, on an event-by-event basis~\cite{Gardim:2020sma,Mu:2025gtr}. 
We denote by $\delta f$ the fluctuation of a quantity $f$ around its mean value at fixed $b$, which we denote simply by $\langle f\rangle$ all the way through Sec.~\ref{s:gaussian}: 
\begin{equation}
\label{deltaf}
\delta f\equiv f-\langle f\rangle. 
\end{equation}
As we shall see in Sec.~\ref{s:fits}, the relative standard deviations of $S$ and $R^2$ are only $3\,\%$ in Pb+Pb collisions at $b=0$ at LHC energies. 
We thus expand to first order in the fluctuations. 
Differentiating Eq.~(\ref{effhydro}), we obtain 
\begin{align}
\label{smallfluct}
\frac{\delta N}{\langle N\rangle}&=\frac{\delta S}{\langle S\rangle}\nonumber\\
\frac{\delta [p_T]}{\langle [p_T]\rangle}&=c_s^2(T_{\rm eff}) \left( \frac{\delta S}{\langle S\rangle}-\frac{3}{2}\frac{\delta R^2}{\langle R^2\rangle}  \right),
\end{align}
where $c_s(T)$ is the speed of sound  at temperature $T$, defined by $c_s^2=d\ln T/d\ln s$ for a baryonless fluid~\cite{Ollitrault:2007du}. 
The variation of $T_{\rm eff}$ with centrality is negligible in the centrality range considered in this paper, and we write $c_s^2$ as a shorthand for $c_s^2(T_{\rm eff})$ from now on. 
Eq.~(\ref{smallfluct}) expresses the final-state fluctuations in terms of initial-state fluctuations. 

We now sketch how $c_s$ can be extracted from data on ultra-central collisions. 
We denote by $N_{\rm knee}$ the knee of the distribution of $N$, defined by~\cite{Das:2017ned} $N_{\rm knee}\equiv\langle N\rangle_{b=0}$. 
For values of $N$ significantly above $N_{\rm knee}$, all events have  $b\approx 0$. 
If $R^2$ is uncorrelated with $S$, then $\langle\delta R^2|S\rangle=0$, where $\langle f|S\rangle$ denotes an average value at fixed $S$ and $b$. 
Applying Eq.~(\ref{smallfluct}), one obtains the following expression for the mean value of $[p_T]$ at fixed $N$: 
\begin{equation}
\label{nocorr1ucc}
\langle [p_T]|N\rangle\approx\langle [p_T]\rangle_{b=0} \left(1+ c_s^2
  \frac{N-N_{\rm knee}}{N_{\rm knee}}\right).
\end{equation}
This equation states that $\langle [p_T]|N\rangle$ increases linearly with $N$, which is seen experimentally, and that $c_s^2$ can be extracted from the slope~\cite{Gardim:2019brr,CMS:2024sgx}. 
Eq.~(\ref{smallfluct}) also shows that at fixed $N$, the variance of $[p_T]$ is induced by fluctuations in the size $R^2$~\cite{Broniowski:2009fm,Bozek:2012fw}, a phenomenon referred to as ``size-flow transmutation''~\cite{Bozek:2017elk}. 

Note that Eq.~(\ref{nocorr1ucc}) no longer holds if $S$ and $R^2$ are correlated~\cite{Nijs:2023bzv,Sun:2024zsy}. 
We will argue that this  correlation is the {\it only\/} irreducible source of uncertainty in the determination of $c_s$ from data on $[p_T]$ fluctuations. 
Fortunately, general theoretical arguments~\cite{Zhou:2025bwu}, as well as global theory-to-data comparisons~\cite{Bernhard:2016tnd,Nijs:2020roc,JETSCAPE:2020mzn}, favor initial-state models where $R^2$ and $S$ are uncorrelated at $b=0$~\cite{Zhou:2025bwu}. 

If $R^2$ and $S$ are uncorrelated, on the other hand, the determination of $c_s^2$ from the slope is very precise, as shown by detailed hydrodynamic simulations at $b=0$~\cite{Gardim:2024zvi}. 
Surprisingly, it is much more precise than Eqs.~(\ref{effhydro}) from which it is derived~\cite{Gavassino:2025bts,SoaresRocha:2024drz}, as long as the equation of state does not vary steeply around $T_{\rm eff}$. 
$c_s^2$ obtained from the slope differs from the thermodynamic value by less than $0.01$ for a broad range of colliding energies and freeze-out temperatures (Fig.~2(d) of Ref.~\cite{Gavassino:2025bts}), and for several equations of state (Fig.~3 of Ref.~\cite{Gardim:2024zvi}). 

\section{Correcting for missing particles at low transverse momentum} 
\label{s:v0pt}

We now quantify the biases from the detector acceptance~\cite{Nijs:2023bzv}. 
ATLAS does not detect particles with  $p_T<0.5$~GeV$/c$, which implies that roughly $50\%$ of the charged particles are missing.\footnote{There is also an upper cut at $5$~GeV$/c$, which eliminates a tiny fraction of the multiplicity. These high-$p_T$ particles contribute little to collective flow, so that this cut has in practice little effect.}
We denote by $N_A$ and $[p_{TA}]$ the values of $N$ and $[p_T]$ after this acceptance cut. 

Its effect is not trivial because the distribution of $p_T$ depends on the temperature of the fluid, that is, on $[p_T]$. 
To a good approximation, the shape of the spectrum depends {\it only\/} on $[p_T]$~\cite{Gardim:2019iah}. 
Denoting by $N(p_T)$ the number of particles in a $p_T$ bin, the relative fluctuation of  $N(p_T)$ can be decomposed as the sum of two terms~\cite{Parida:2024ckk}:
\begin{equation}
\label{spectrumfluct}
\frac{\delta N(p_T)}{\langle N(p_T)\rangle}=\frac{\delta N}{\langle N\rangle}+\frac{v_0(p_T)}{v_0}\frac{\delta [p_T]}{\langle [p_T]\rangle}. 
\end{equation}
The first term in the right-hand side represents the change in the overall normalization, while the second term represents the change in the normalized spectrum induced by a small variation of $[p_T]$. 
This is precisely how one defines the observable $v_0(p_T)/v_0$~\cite{Schenke:2020uqq}, which has recently been measured in Pb+Pb collisions at the LHC~\cite{ATLAS:2025ztg,ALICE:2025iud}. 
Upon integration over the interval of $p_T$ covered by the detector, which we denote by $A$, one obtains: 
\begin{align}
\label{totalmultA}
\frac{\delta N_A}{\langle N_A\rangle}&=\frac{\delta N}{\langle N\rangle}+D_A\frac{\delta [p_T]}{\langle [p_T]\rangle}\nonumber\\
\frac{\delta [p_{TA}]}{\langle [p_{TA}]\rangle}&=C_A\frac{\delta [p_T]}{\langle [p_T]\rangle}
\end{align}
where $D_A$ and $C_A$ are dimensionless correction factors defined by  
\begin{align}
\label{defDA}
D_A &\equiv \frac{\int_A \frac{v_0(p_T)}{v_0} \langle N(p_T)\rangle}{\int_A \langle N(p_T)\rangle}\nonumber\\
C_A &\equiv \frac{\int_A \frac{v_0(p_T)}{v_0} (p_T-\langle p_{TA})\rangle \langle N(p_T)\rangle}{\int_A p_T\langle N(p_T)\rangle}. 
\end{align}
$C_A$ has been introduced in Ref.~\cite{Parida:2024ckk}, while $D_A$ is introduced here for the first time. 

\begin{figure}[ht]
\includegraphics[width=\linewidth]{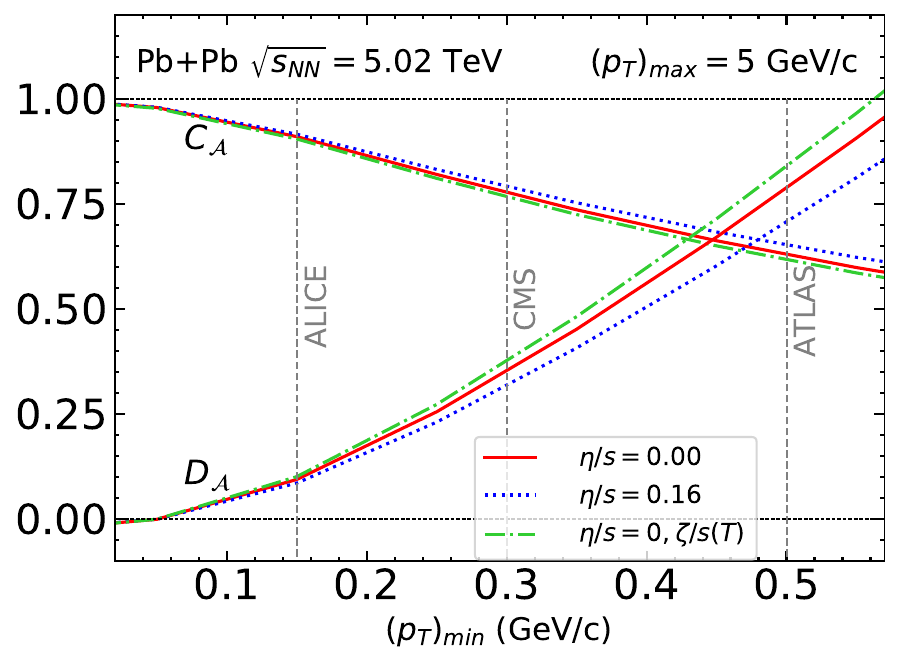}
    \caption{
      Variation of the acceptance factors $C_A$  and $D_A$, defined by Eq.~(\ref{defDA}), as a function of the lower $p_T$ cut. 
      We indicate by vertical lines the cuts implemented by ALICE~\cite{ALICE:2025rtg}, ATLAS~\cite{ATLAS:2024jvf} and CMS~\cite{CMS:2024sgx}. 
      Our results are obtained from hydrodynamic simulations with smooth initial conditions~\cite{Parida:2024ckk}. 
      We run ideal hydrodynamics, then implement separately a constant shear viscosity over entropy $\eta/s$, and the bulk viscosity from the Duke parametrization (Table 4 of Ref.\cite{Moreland:2018gsh}).
      }
     \label{fig:acceptance}
\end{figure}
Both $D_A$ and $C_A$ are to a good approximation independent of centrality, like $v_0(p_T)/v_0$.
Fig.~\ref{fig:acceptance} displays their variation, calculated in hydrodynamics,  as a function of the lower $p_T$ cut.
When the lower $p_T$ cut goes to $0$,  $D_A\to 0$ and $C_A\to 1$.
In this limit, $\delta N_A$ and $\delta [p_{TA}]$ coincide with $\delta N$ and $\delta [p_T]$, as they should. 
Values of $D_A$ and $C_A$ have small sensitivity to initial-state fluctuations~\cite{Parida:2024ckk}, which we therefore ignore, but depend significantly on transport coefficients.
These coefficients are poorly constrained from data~\cite{JETSCAPE:2020mzn}, and little is known about the resulting correction to the momentum distribution at freeze-out~\cite{JETSCAPE:2020mzn,Dusling:2011fd}. 
Therefore, we calibrate the values of $C_A$ and $D_A$ against ATLAS data.  
Specifically, we use data on the increase of the variance of $[p_T]$, as the upper $p_T$ cut evolves from 2 to 5 GeV$/c$~\cite{ATLAS:2022dov}. 
Fig.~4 of Ref.~\cite{Parida:2024ckk} shows that this increase is overestimated by ideal hydrodynamics,  and underestimated with shear viscosity $\eta/s=0.16$, so that we take the full red lines and dotted blue lines in Fig.~\ref{fig:acceptance} as the bounds. 
With the lower $p_T$ cut of ATLAS, we obtain $C_A=0.64\pm 0.01$ and $D_A=0.75\pm 0.05$.

Using Eqs.~(\ref{smallfluct}) and (\ref{totalmultA}), one derives the following relation between final-state fluctuations $\delta N_A$ and $\delta [p_{TA}]$ and initial-state fluctuations $\delta S$ and $\delta R^2$, which we write in matrix form:  
\begin{equation}
\label{linearresponse}
\begin{pmatrix}
\frac{\delta N_A}{\langle N_A\rangle}\\
\frac{\delta [p_{TA}]}{\langle [p_{TA}]\rangle}
\end{pmatrix}
=
\begin{pmatrix}
1+D_Ac_s^2 &-\frac{3}{2}D_Ac_s^2\\
C_Ac_s^2 & -\frac{3}{2}C_Ac_s^2
\end{pmatrix}
\begin{pmatrix}
\frac{\delta S}{\langle S\rangle}\\
\frac{\delta R^2}{\langle R^2\rangle}
\end{pmatrix}.
\end{equation}
Comparison with Eq.~(\ref{smallfluct}) shows the changes induced by the detector acceptance: 
$C_A$ amounts to renormalizing the value of $c_s^2$, while $D_A$ introduces a new coupling between $\delta N_A$ and $\delta R^2$~\cite{Nijs:2023bzv}. 
Due to this coupling, Eq.~(\ref{nocorr1ucc}) no longer holds, and one can no longer extract $c_s^2$ from $\langle [p_{TA}]\rangle$ data alone. 
This information must be combined with data on the variance of $[p_{TA}]$, as we now explain.

\section{Modeling fluctuations at fixed $b$}
\label{s:gaussian}

Fluctuations at fixed $b$ are Gaussian to a good approximation~\cite{Samanta:2023amp}: 
The joint distribution of $\delta S$ and $\delta R^2$ is a centered 2-dimensional Gaussian, and so is that of $\delta N_A$ and $\delta [p_{TA}]$, since they are related by a linear mapping (Eq.~(\ref{linearresponse})). 
The joint distribution of $N_A$ and $[p_{TA}]$ is a non-centered 2-dimensional Gaussian.  
Experimentally, the mean and variance of $[p_{TA}]$ are evaluated at fixed multiplicity, that is, fixed $N_A$. 
For a 2-dimensional Gaussian, the conditional mean and variance of one of the variables when the other is fixed are given by: 
\begin{align}
  \langle [p_{TA}]|N_A\rangle &=\langle [p_{TA}]\rangle+\frac{\langle \delta [p_{TA}]\delta N_A\rangle}{\langle(\delta N_A)^2\rangle}\delta N_A\label{gaussianmoment1}\\
  {\rm Var}([p_{TA}]|N_A) &= \langle(\delta [p_{TA}])^2\rangle-
  \frac{\langle \delta [p_{TA}]\delta N_A\rangle^2}{\langle(\delta N_A)^2\rangle}.
  \label{gaussianmoment2}
\end{align}
%The mean value of $[p_{TA}]$ is linear in $N_A$, while the variance is independent of $N_A$. 
In the case of an ideal acceptance, using Eqs.~(\ref{smallfluct}) and (\ref{gaussianmoment1}), one recovers Eq.~(\ref{nocorr1ucc}) if $S$ and $R^2$ are uncorrelated. 
With acceptance corrections taken into account, however, Eq.~(\ref{linearresponse}) shows that the right-hand side of Eq.~(\ref{gaussianmoment1}) now involves the variance of $R^2$, which is unknown. 
This variance can be inferred from that of $[p_{TA}]$ (see the discussion  below Eq.~(\ref{nocorr1ucc})). 
Thus we need to combine information from the variance and from the mean of $[p_{TA}]$ in order to infer the value of $c_s^2$.

A 2-dimensional Gaussian distribution has 5 parameters: 2 mean values, 2 variances, and 1 covariance, which are all smooth functions of $b$, for which we choose generic, theory-agnostic parametrizations  (Appendix~\ref{s:fittingfunctions}). 
Our approach is data driven, and we infer as much information as possible on these 5 functions from data. 
Using the distribution of $N_{ch}$, we infer the mean value of $N_A$ as a function of $b$, as well as the variance of $N_A$ at $b=0$~\cite{Das:2017ned} (Sec.~\ref{s:geometry}).
The mean of $[p_{TA}]$ and the variance at fixed $N_A$ (right-hand side of Eq.~(\ref{gaussianmoment2})) as a function of $b$ are inferred from data on the mean and variance of $[p_{TA}]$ (Sec.~\ref{s:fits}).

There are two pieces of information that cannot be obtained from data: 
The $b$-dependence of the variance of $N_A$, and the covariance between $N_A$ and $[p_{TA}]$.
Here, we need to resort to initial-state models. 
In particular, we assume that the covariance of $S$ and $R^2$ vanishes at $b=0$. 
We relate the final-state covariance between $N_A$ and $[p_{TA}]$, which appears in Eq.~(\ref{gaussianmoment1}),  to this initial-state covariance, using the linear mapping (\ref{linearresponse}).  
The algebra is detailed in Appendix~\ref{s:mappingdetails}. 
The uncertainties from initial-state models are evaluated in Sec.~\ref{s:fits}. 

\section{Impact parameter fluctuations}
\label{s:geometry}

We have derived the expressions of the mean and variance of $[p_{TA}]$ at fixed $b$ and $N_A$. 
In order to compare with experiment, they must be averaged over $b$ at fixed $N_{ch}$. 
We now explain how to average over $b$ at fixed $N_A$, and the average over $N_A$ at fixed $N_{ch}$ will be studied in Sec.~\ref{s:poisson}. 

\begin{figure}[t]
\includegraphics[width=\linewidth]{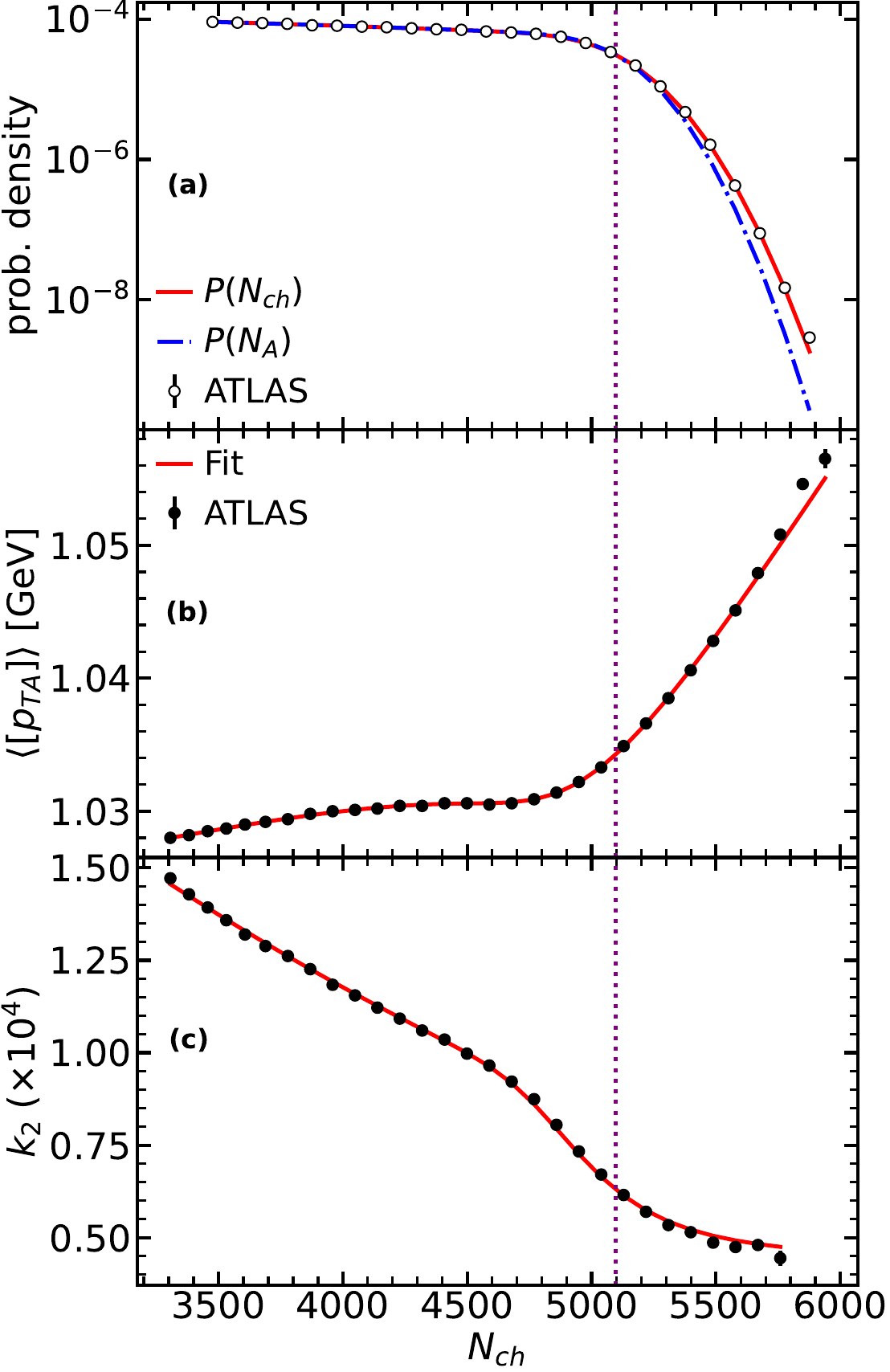}
    \caption{
    Symbols are ATLAS data~\cite{ATLAS:2024jvf}, lines are fits. 
    (a) Probability of the charged-particle multiplicity $N_{ch}$. We only display results for $N_{ch}\gtrsim 3500$, corresponding to 10\% of the total number of events. 
    The full line is a fit assuming Gaussian fluctuations at fixed $b$. The dash-dotted line is the distribution of $N_A$ after unfolding Poisson fluctuations (Sec.~\ref{s:geometry}). 
    (b) Event-averaged transverse momentum per particle $\langle [p_{TA}]\rangle$. 
    (c) Relative variance  $k_2\equiv {\rm Var}([p_{TA}])/\langle [p_{TA}]\rangle^2$.  
}      
     \label{fig:ATLASfits}
\end{figure}

The first step is to reconstruct the distribution of $N_{ch}$ at fixed $b$. 
This is a standard procedure~\cite{INDRA:2020kyj,Parfenov:2021ipw,INDRA:2023myq,CMS:2024sgx,ALICE:2025rtg,Idrisov:2025kog,Roubertie:2025qps}. 
Assuming that the distribution of $N_{ch}$ at fixed $b$ is Gaussian~\cite{Das:2017ned}, one reconstructs the variation of the mean with $b$, $\langle N_{ch}|b\rangle$, by fitting the measured probability $P(N_{ch})$  as a superposition of Gaussians (full line in Fig.~\ref{fig:ATLASfits} (a)). 
The knee of the distribution is $N_{\rm knee}\equiv \langle N_{ch}|b=0\rangle\approx 5100$. 

The variance of $N_{ch}$ at $b=0$, which we denote by $\sigma_{N_{ch}}^2$, determines the tail of $P(N_{ch})$ and is also reconstructed  precisely~\cite{Das:2017ned}. 
We obtain $\sigma_{N_{ch}}\approx 192$. 
The relative variance  is $\sigma_{N_{ch}}^2/N_{\rm knee}^2\approx 1.42\times 10^{-3}$.\footnote{
Note that it is significantly smaller than the value $2.0\times 10^{-3}$ obtained~\cite{Yousefnia:2021cup} from earlier ATLAS data~\cite{ATLAS:2019peb}. 
This reduction might be due to a tighter rejection of pile-up events in the new analysis.}
The  $b$ dependence of the variance cannot be determined from data~\cite{Das:2017ned}, as mentioned in Sec.~\ref{s:gaussian}. 
The parametrization of this $b$ dependence is specified in Appendix~\ref{s:fittingfunctions}. 

We now explain how to relate the discrete hadron multiplicity $N_{ch}$, which is measured in experiment, to the continuous $N_A$ calculated in hydrodynamics. 
In hydrodynamics, hadrons are sampled independently from the elements of the freeze-out hypersurface~\cite{Cooper:1974mv,Gale:2013da}, so that a single hydrodynamic calculation  describes an ensemble of events, rather than a single event. 
For a given $N_A$, the number of particles $N_{trk}$ seen in the detector approximately follows a Poisson distribution, which implies 
\begin{equation}
\label{varpoisson}
{\rm Var}(N_{trk}|N_A)=\langle N_{trk}|N_A\rangle.
\end{equation}
In the ATLAS analysis, the values of  $N_{ch}$ are corrected for the reconstruction efficiency: 
\begin{equation}
\label{efficiency}
N_{ch}=\frac{1}{\varepsilon}N_{trk},
\end{equation} 
where  $\varepsilon\approx 0.73$  in central collisions~\cite{ATLAS:2024jvf}.\footnote{Note that the reconstruction efficiency depends on transverse momentum $p_T$ and pseudorapidity $\eta$. We use the value averaged over the acceptance.} 
We normalize $N_A$ so that it corresponds to the mean value of $N_{ch}$: 
\begin{equation}
\label{srenorm}
\langle N_{ch}|N_A\rangle=N_A.
\end{equation} 
Eqs.~(\ref{varpoisson}) and (\ref{efficiency}) then give: 
\begin{equation}
\label{varpoissoncorr}
{\rm Var}(N_{ch}|N_A)=\frac{1}{\varepsilon}N_A.
\end{equation}
In the ATLAS experiment, the number of tracks is $N_{trk}\approx 3720$ in a collision at $b=0$~\cite{ATLAS:2022dov}, so that the relative standard deviation is
$3720^{-1/2}\approx 1.6\%$.  
In this regime of small fluctuations, the Poisson distribution reduces to a Gaussian: 
\begin{equation}
\label{poissongauss}
P(N_{ch}|N_A)\approx  \frac {1}{\sqrt{2\pi N_A/\varepsilon}}\exp\left(-\frac{(N_{ch}-N_A)^2}{2N_A/\varepsilon}\right). 
\end{equation}

Once the distribution of $N_{ch}$ at fixed $b$ is known, it is straightforward to reconstruct the distribution of $N_A$ at fixed $b$. 
Decomposing an average over $N_{ch}$ at fixed $b$ as an average over $N_{ch}$ at fixed $N_A$ and $b$, followed by an average over $N_A$, one obtains:  
\begin{align}
\label{decpoisson}
 \langle N_{ch}|b\rangle &=\langle N_A|b\rangle\nonumber\\
{\rm Var}(N_{ch}|b) &=  \langle {\rm Var}(N_{ch}|N_A)|b\rangle+ {\rm Var}(N_A|b)\nonumber\\ &\equiv \frac{1}{\varepsilon}\langle N_{ch}|b\rangle +\sigma_{N_A}(b)^2, 
\end{align}
where we have used Eq.~(\ref{varpoissoncorr}).
Thus the variance of $N_{ch}$ is decomposed into a statistical part, corresponding to Poisson fluctuations, and a dynamical part, corresponding to fluctuations of $N_A$. 
Eq.~(\ref{decpoisson}) allows us to evaluate the width ${\sigma_{N_A}}(b)$ of dynamical fluctuations.  
They  originate from initial conditions, not from the hadronization process~\cite{Yousefnia:2021cup}. 
The distribution of $N_A$ at fixed $b$ is Gaussian, like that of $N_{ch}$, but with a reduced width. 
At $b=0$, statistical and dynamical fluctuations represent 19\,\% and 81\,\% of the variance~\cite{Yousefnia:2021cup}. 

The distribution of $N_A$ is obtained after averaging over $b$. 
It is displayed in Fig.~\ref{fig:ATLASfits} (a) and essentially coincides with the distribution of $N_{ch}$, except in the tail, which is narrower because only the dynamical variance enters. 
Once the probability distribution of $N_A$ at fixed $b$ is known, that of $b$ at fixed $N_A$ is given by Bayes' theorem~\cite{Das:2017ned}. 

%\subsection{Effect of hadronization on the extraction of $c_s^2$}
%\label{s:poissoncs}

\section{Deblurring hadronization}
\label{s:poisson}

The observables of interest are the moments of $[p_{TA}]$. 
Experimentally, they are measured for events with the same $N_{ch}$, while in our hydrodynamic picture, they are calculated for events with the same $N_A$. 
Fluctuations of $N_{ch}$ relative to $N_A$, due to the hadronization process, act as a noise which blurs the smooth hydrodynamic image. 
This noise must be modeled in order to compare with experimental data, which are at fixed $N_{ch}$: 
Specifically, we need to average the moments  of $[p_{TA}]$ over $N_A$ at fixed $N_{ch}$. 

The first step is to determine the distribution of $N_A$ for a given $N_{ch}$, which is given by Bayes' theorem: 
\begin{equation}
\label{bayes}
P(N_A|N_{ch})= \frac{ P(N_{ch}|N_A) P(N_A)}{P(N_{ch})}.
\end{equation}
$P(N_{ch}|N_A)$ is a narrow Gaussian (\ref{poissongauss}), of width $\sigma\approx\sqrt{N_{ch}/\varepsilon}$, while $P(N_A)$ is a slowly-varying function. 
The product is approximately a Gaussian with the same width, but whose mean is slightly displaced according to:\footnote{This result is exact if the distribution of $N_{ch}$ is Gaussian. In practice, we use the following analytic formula, which is a very good approximation~\cite{Gardim:2019brr}:  $$\bar N_A= N_{ch}-\sqrt{\frac{2}{\pi}}\frac{\sigma^2}{\sigma_{N_{ch}}}\frac{\exp\left(-\frac{(N_{ch}-N_{\rm knee})^2}{2\sigma_{N_{ch}}^2}\right)}{{\rm erfc}\left(\frac{N_{ch}-N_{\rm knee}}{\sqrt{2}\sigma_{N_{ch}}}\right)}.
$$ 
 }

\begin{equation}
\label{shift}
\bar N_A\equiv \langle N_A|N_{ch}\rangle\approx N_{ch}+\sigma^2\frac{d}{dN_{ch}} \ln P(N_{ch}). 
\end{equation}
This shift is large in ultracentral collisions, where $P(N_{ch})$ steeply decreases. 
Above the knee, $P(N_{ch})$ is approximately Gaussian: 
\begin{equation}
P(N_{ch})\propto \exp\left(-\frac{(N_{ch}-N_{\rm knee})^2}{2 \sigma_{N_{ch}}^2}\right). 
\end{equation}
Inserting into Eq.~(\ref{shift}), with $\sigma^2\approx N_{\rm knee}/\varepsilon$, one obtains
\begin{equation}
\label{renorm}
\bar N_A-N_{\rm knee}= \left(1-\frac{N_{\rm knee}}{\varepsilon\sigma_{N_{ch}}^2}\right)\left(N_{ch}-N_{\rm knee}\right). 
\end{equation}
The consequence on the determination of $c_s^2$ can be easily derived for an ideal detector, where $N_A=N$. 
Inserting Eq.~(\ref{renorm}) into  Eq.~(\ref{nocorr1ucc}), one sees that the effect of $c_s^2$ is renormalized by a factor   $\approx 0.81$.\footnote{This effect was referred to as a self-correlation in Ref.~\cite{Nijs:2023bzv}. The correction proposed in Eq.~(5) of Ref.~\cite{Gardim:2024zvi} is wrong, the exponent should be $-1$, not $-\frac{1}{2}$. This mistake was pointed out to us by Matt Luzum.} 
Therefore, $c_s^2$ is underestimated by $19\%$ if one does not properly unfold Poisson fluctuations.
With the detector acceptance taken into account, the correction is even larger,  $27\%$. 
It is also essential to know the reconstruction efficiency $\varepsilon$ implemented in the analysis. 
If one forgets this correction by setting $\varepsilon=1$, the estimated $c_s^2$ is reduced by $10\%$. 

The second step of the deblurring is to average a slowly-varying function  $f(N_A)$  (in our case, a moment of $[p_{TA}]$ at fixed $N_A$) over $N_A$ at fixed $N_{ch}$. This average is the integral: 
\begin{equation}
I\equiv  \int f(N_A) P(N_A|N_{ch})dN_A. 
\end{equation}
One can derive successive approximations of this integral by requiring that the result is exact if  $P(N_A|N_{ch})$ is Gaussian, and if $f(N_A)$ is a polynomial of increasing degree. 
The lowest-order approximations are: 
\begin{align}
\label{NLO}
{\rm LO}:\ &  I\approx f(\bar N_A)  \nonumber\\
{\rm NLO}:\ &  I\approx  \frac{1}{2}f(\bar N_A-\sigma)+ \frac{1}{2}f(\bar N_A+\sigma)\\
{\rm NNLO}:\ & I\approx  \frac{1}{4}f(\bar N_A-\sigma\sqrt{2})+ \frac{1}{2}f(\bar N_A)+ \frac{1}{4}f(\bar N_A+\sigma\sqrt{2}), \nonumber
\end{align}
which are exact for polynomials of degree 1 (leading order LO), 3 (next-to-leading order NLO) and 5 (next-to-next-to-leading order NNLO).
We find sizable differences between NLO and LO results, but NNLO and NLO are essentially identical, so that we choose the NLO formulas for the results presented in Sec.~\ref{s:fits}. 

\section{Fitting ATLAS data}
\label{s:fits}

The ATLAS analysis~\cite{ATLAS:2024jvf} classifies events according to the value of $N_{ch}$. 
This is a major difference compared to the pioneering analysis by the  CMS Collaboration~\cite{CMS:2024sgx}, which uses a separate centrality classifier. 
Such a third-party centrality classifier has in general non-trivial correlations with $N_{ch}$ and $[p_{TA}]$.\footnote{ATLAS also defines centrality according to a different observable, which is the energy deposited in forward calorimeters. In practice, however, this has no effect on the analysis carried out in this paper, since events are binned according to the value of $N_{ch}$, not to the value of centrality.}
They bias the quantity of interest, which is the correlation between $[p_{TA}]$ and $N_{ch}$~\cite{Nijs:2023bzv}. 
This has been shown in detail by the ALICE Collaboration~\cite{ALICE:2025rtg}, which has compared nine different estimators. 
The motivation for using a separate centrality classifier was to avoid the ``self correlation" with the charged multiplicity used for the analysis, which induces another bias. 
This self-correlation is taken care of by the unfolding of statistical fluctuations carried out in Sec.~\ref{s:poisson}, and is no longer a problem. 

ATLAS measures the mean value and the relative variance $k_2$ of $[p_{TA}]$ as a function of $N_{ch}$ 
(Fig.~\ref{fig:ATLASfits} (b) and (c)). 
They subtract the contribution of self-correlations from the variance, and only include the contribution of two-particle correlations. 
Therefore, $k_2$  measures the dynamical fluctuations of $[p_{TA}]$~\cite{STAR:2005vxr} and would vanish for uncorrelated particles. 
Note, however, that no rapidity gap between the two particles is implemented, so that there could be a contribution to the variance from short-range nonflow correlations~\cite{Parida:2024ckk}. 
 We assume that they are negligible. 

We fit our model to these data by going through the following steps. 
We start by evaluating the first two moments of $[p_{TA}]$ at fixed $N_A$ and $b$. 
The moment of order 1 is given by Eq.~(\ref{gaussianmoment1}), while the moment of order 2 is decomposed as  
\begin{equation}
\label{moment2}
\langle [p_{TA}]^2 |N_A,b\rangle=\langle [p_{TA}]|N_A,b\rangle^2+{\rm Var}( [p_{TA}]|N_A,b),
\end{equation}
where the two terms are given by Eqs.~(\ref{gaussianmoment1}) and (\ref{gaussianmoment2}).
We then average these two moments over $b$ at fixed $N_A$ as explained in Sec.~\ref{s:geometry}, and finally over $N_A$ at fixed $N_{ch}$ as explained in Sec.~\ref{s:poisson}. 
We evaluate the relative variance as $k_2=\langle [p_T]^2|N_{ch}\rangle/\langle [p_T]|N_{ch}\rangle^2-1$. 

Our fits are displayed as solid lines in Fig.~\ref{fig:ATLASfits} (b) and (c).
In addition to $c_s^2$, the fit returns the values of the mean $\langle [p_{TA}]\rangle$, and the variance of $[p_{TA}]$ at fixed $N_A$, as a function of $b$.
We evaluate the error bars on fit parameters as follows. 
We vary the range of the fit, and more specifically, the minimum value of $N_{ch}$, from 4500 down to 2500, so that the fraction of events included in the fit varies between $\approx 3\%$ and $\approx 20\%$. 
In addition, we vary the parameters which are taken from initial-state models:
The centrality dependence of the relative variance of $N_{ch}$, and that of the covariance between $S$ and $R^2$. 

For $c_s^2$, there is one irreducible source of uncertainty, as mentioned in Sec.~\ref{s:hydropicture}, which is the value of the correlation between $S$ and $R^2$ at $b=0$. 
Our default assumption is that it vanishes, as favored by general arguments~\cite{Zhou:2025bwu}. 
If it does not, the relative change in the fitted $c_s^2$ is $\approx 1.5\,\rho$, where $\rho$ is the Pearson correlation between $S$ and $R^2$ at $b=0$. 
%%%JYO: I write $1.5\rho$, which is what one derives analytically for a perfect detector. For ATLAS, the effect is somewhat larger, $1.8\rho$, but on the other hand it will be closer to $1.5$ is the analysis is repeated by other collaborations, so I prefer to mention this simple general result, at the expense of precision. 
Global theory-to-data comparisons~\cite{Bernhard:2016tnd,Nijs:2020roc,JETSCAPE:2020mzn} could easily put bounds on $\rho$, but this is beyond the scope of the present work.

The largest error on $c_s^2$ comes from varying the range of the fit.
The next-to-largest errors are from the $C_A$ and $D_A$ coefficients, which are evaluated in hydrodynamics.
Next comes the error from the $b$ dependence of the variance of $N_{ch}$.
The smallest source of error is the correlation between $S$ and $R^2$ for $b>0$, which is predicted to be positive (Appendix~\ref{s:trento}), but this has a very small effect on $c_s^2$. 
Putting together all errors, we finally obtain:
\begin{equation}
  \label{csfit}
  c_s^2=0.246\pm 0.008.
\end{equation}

There are other interesting results from the fit.
The impact parameter dependence of $\langle [p_{TA}]\rangle$ is tiny but not trivial: 
It slightly increases up to $\approx 4\,\%$ centrality and then decreases. 
Interestingly, this trend is also predicted by initial-state models, supplemented by the effective hydrodynamic description (\ref{effhydro}), as shown in Appendix~\ref{s:trento}. 

We also obtain non-trivial information about initial-state fluctuations. 
We determine the relative standard deviations of $S$ and $R^2$ at $b=0$ using the reverse-engineering formulas derived in Appendix~\ref{s:mappingdetails}:  
\begin{align}
  \label{reverseengineering}
  \frac{\langle (\delta S)^2\rangle^{1/2}}{\langle S\rangle} &= 2.77\pm 0.05\,\% \nonumber\\
  \frac{\langle (\delta R^2)^2\rangle^{1/2}}{\langle R^2\rangle} &= 3.09\pm 0.15\,\%  .
\end{align} 
Note that the relative standard deviation of $S$ is smaller than that of $N_{ch}$ by $\approx 26\%$ due to Poisson fluctuations and to the low-$p_T$ cut. 
Note also that the relative standard deviation of $R^2$ is smaller than in standard initial-state calculations (Appendix~\ref{s:trento}) which give a value $\approx 4\, \%$.
%Here, the largest uncertainty comes from the values of the acceptance coefficients $D_A$ and $C_A$ (Eq.~(\ref{defDA})) which come from a hydrodynamic calculation. 

\section{Comparison with first-principles calculations}
\label{s:lattice}

\begin{figure}[ht]
   \includegraphics[width=\linewidth]{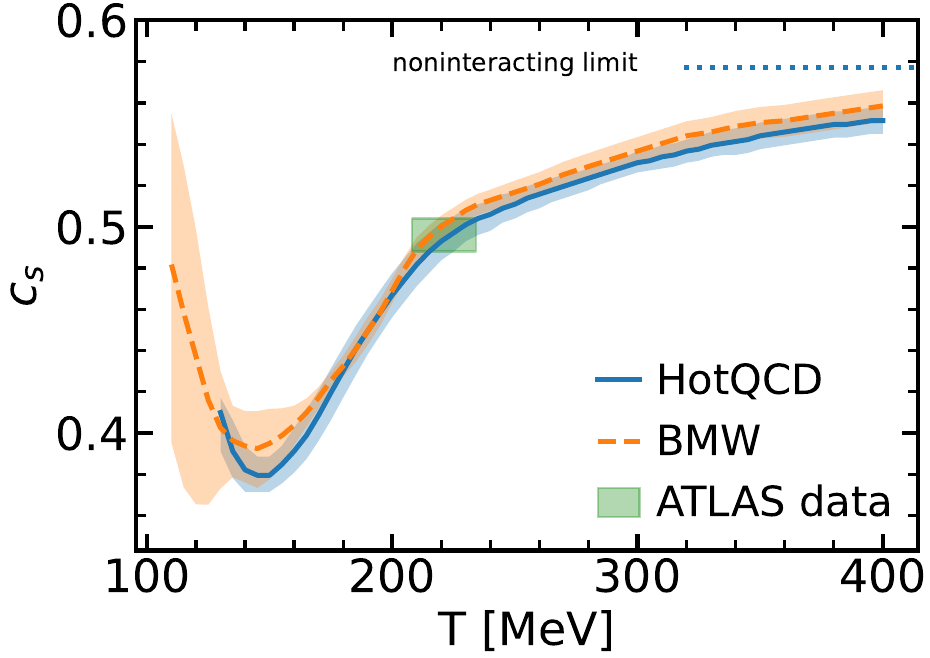}
    \caption{Comparison between the value of $c_s$ obtained by fitting ATLAS data~\cite{ATLAS:2024jvf} and lattice QCD results
   by the BMW~\cite{Borsanyi:2013bia} and HotQCD~\cite{HotQCD:2014kol} Collaborations. 
}      
     \label{fig:lattice}
\end{figure}

The value of $c_s$ obtained by fitting ATLAS data is very robust against the details of the calculation. 
There is a much larger uncertainty on the temperature $T_{\rm eff}$ at which $c_s$ is evaluated, than on the value of $c_s$ itself.
There are theoretical uncertainties on the ratio $\langle p_T\rangle/T_{\rm eff}$, as well as experimental uncertainties on $\langle p_T\rangle$. 
Let us start with the latter. 
The ALICE Collaboration finds $\langle p_T\rangle=681\pm 9$~MeV$/c$~\cite{ALICE:2018hza}, while CMS gives $\langle p_T\rangle=658\pm 25$~MeV$/c$~\cite{CMS:2024sgx}.
As for the ratio $\langle p_T\rangle/T_{\rm eff}$, it was originally found to be $3.07$ with a smooth pure hydrodynamic calculation~\cite{Gardim:2019xjs}, but the ratio was lowered to $3.00\pm 0.05$ after including a hadronic afterburner~\cite{Gardim:2024zvi}.
When including initial-state fluctuations, but without an afterburner, the ratio goes down from $3.07$ to $3.03$~\cite{Gardim:2020sma}.
If the effect of fluctuations reduces the ratio by a similar amount when a hadronic afterburner is implemented, the ratio would go down to $2.96$. 
Putting all these uncertainties together, we obtain the conservative estimate
\begin{equation}
  \label{teffestimate}
  T_{\rm eff}=221\pm 13~{\rm MeV}. 
\end{equation}
This could be refined by carrying out hydrodynamic simulations at $b=0$ with initial-state fluctuations and a hadronic afterburner. 
Our results are in perfect agreement with lattice QCD calculations, as shown in Fig.~\ref{fig:lattice}.

\section{Perspectives}
\label{s:conclusions}

We have extracted a robust estimate of the speed of sound in the quark-gluon plasma from ultra-central collision data, using a simplified hydrodynamic description, supplemented by an accurate Bayesian reconstruction of impact parameter fluctuations. 
We correct for all the known biases.
Biases due to the limited $p_T$ coverage and to statistical fluctuations at hadronization lead to corrections on $c_s^2$ which are large, typically $30\,\%$.
It is quite remarkable that after correcting for these biases, the value of $c_s$ inferred from heavy-ion data is in perfect agreement with predictions from lattice QCD.

%%The weakness of our approach is the effective hydrodynamic description on which it relies. 
%%More sophisticated event-by-event hydrodynamic calculations are necessary in order to improve over this description and/or evaluate the resulting error, which we are unable to quantify. 
%%On the other hand, the strength of our approach is that impact parameter fluctuations, which play a crucial role, are precisely evaluated in a data-driven way, which is not implemented in any of the state-of-the-art models~\cite{Bernhard:2016tnd,Nijs:2020roc,JETSCAPE:2020mzn}. 

Our analysis of ATLAS data can easily be repeated by other collaborations, CMS and ALICE at the LHC, and STAR at RHIC. 
Their lower $p_T$ cuts are significantly lower than that of ATLAS, which reduces the theoretical uncertainties from acceptance cuts (through the coefficients $D_A$ and $C_A$). 
In particular, our procedure could be applied to ALICE data on $[p_T]$  with centrality classifier I~\cite{ALICE:2025rtg}, where events are binned directly according to $N_{ch}$, as in the ATLAS analysis. 
Some information is still missing, which prevents us from fitting ALICE data in the same way as ATLAS data: 
The normalizations of the variance of $[p_T]$ and of the charged multiplicity  are not provided, and the reconstruction efficiency is not specified.
Therefore, the best we can do at this stage is to guess how the value reported in Fig.~4 of Ref.~\cite{ALICE:2025rtg}, $c_s^2\approx 0.1369\pm 0.0017$, would be modified after taking into account the various corrections implemented in this paper: 
\begin{itemize}
\item
ALICE extracts $c_s^2$ using the fit formula of Ref.~\cite{Gardim:2019brr}, which assumes that $\langle [p_T]|c\rangle$ is independent of centrality $c$, i.e., a flat baseline below the knee. 
We relax this assumption and fit the centrality dependence of $\langle [p_T]|c\rangle$ (Fig.~\ref{fig:trento}). 
One sees in Fig.~4 of Ref.~\cite{ALICE:2025rtg} that the slope below the knee is $\approx -0.02$. The value of $c_s^2$ is the slope relative to this baseline, and one must add $\approx 0.02$ to the value above. 
\item
The low-$p_T$ acceptance cut (Sec.~\ref{s:v0pt})  increases $c_s^2$ by an extra factor $1/C_A\sim 1.11$ (Fig.~\ref{fig:acceptance}), giving $c_s^2\approx 0.174$.
\item 
The largest correction is from Poisson fluctuations at hadronization (Eq.~(\ref{renorm})). 
For ATLAS, $c_s^2$ increases by $1/0.81$ after taking into account this effect. Now, the relative increase is inversely proportional to the multiplicity $N_{trk}$ seen in the detector. 
Using $dN_{ch}/d\eta\approx 2035$~\cite{ALICE:2015juo} and assuming a  reconstruction efficiency $\varepsilon\approx 0.75$~\cite{Arslandok:2022dyb}, we evaluate  $N_{trk}\approx 2440$ in the $|\eta|<0.8$ window for ALICE, as opposed to $N_{trk}\approx 3720$ for ATLAS, implying a increase of $c_s^2$ by $\approx 36\%$. This yields $c_s^2\approx 0.237$, quite close to our estimate (\ref{csfit}). 
\end{itemize}
We conjecture that the residual discrepancy will be resolved after carrying out a full combined fit of ALICE data.

Our procedure can be improved. 
We have assumed that the multiplicity at freeze-out follows a Poisson distribution, which amounts to neglecting correlations between outgoing hadrons. 
Correlations from resonance decays (most notably, $\eta$ and $\rho$ decay in which the final state contains two charged pions) will typically increase the variance by an amount which is likely modest, but should be evaluated as precisely as possible. 
Similarly, nonflow correlations may increase the variance of $[p_T]$, and our model does not capture this increase.  
If the analysis is repeated with a lower $p_T$ cut (by STAR or ALICE), $c_s^2$ will be inferred essentially from the mean, and the uncertainty from nonflow effects will correspondingly be reduced. 

Finally, the skewness of $[p_T]$ fluctuations, which has also been precisely measured by ALICE and ATLAS, can be studied along the same lines. 
This observable receives large contributions from impact parameter fluctuations~\cite{Samanta:2023kfk}, so that it can probably shed light on how the centrality resolution evolves as a function of the collision multiplicity, which is not known at present. 
This study is left for future work. 

\begin{acknowledgments}
We thank Somadutta Bhatta for numerous discussions about the ATLAS results and comments on the draft, Rupam Samanta for discussions about acceptance cuts, and Jean-Paul Blaizot for comments on the draft. 
JYO thanks Matt Luzum for an illuminating discussion about Poisson fluctuations. 
M. Alqahtani acknowledges the support of the Research Mobility Program of the French Embassy in Riyadh. T. Parida acknowledges support from the AGH University of Krakow and the Polish National Science Centre grant:2023/51/B/ST2/01625.
\end{acknowledgments}

\appendix

\section{Initial-state fluctuations versus final-state fluctuations}
\label{s:mappingdetails}

In this Appendix, we derive the expressions relating the covariance matrices of initial-state fluctuations (of $S$ and $R^2$) and final-state fluctuations (of $N_A$ and $[p_{TA}]$) at fixed $b$. 
To simplify expressions, we rewrite Eq.~(\ref{linearresponse}) in matrix form: 
\begin{equation}
\label{matrixform}
Y=M X,
\end{equation}
where $X$ and $Y$ are two-component vectors quantifying the relative initial-state and final-state fluctuations. 
We denote by $C_{ij}$ the relative covariance matrix in the initial state, and by $\Sigma_{ij}$ that in the final state. 
In matrix notation, they are defined by 
\begin{align}
\label{covariances}
C&\equiv\langle XX^T\rangle\nonumber\\
\Sigma&\equiv\langle YY^T\rangle. 
\end{align}
We denote by $|\Sigma|$ the determinant of $\Sigma$. 
With these notations, Eqs.~(\ref{gaussianmoment1})-(\ref{gaussianmoment2}) can be rewritten as: 
\begin{align}
  \langle [p_{TA}]|N_A\rangle &=\left(1+ \frac{\Sigma_{12}}{\Sigma_{11}}\delta N_A\right)\langle [p_{TA}]\rangle\nonumber\\
  {\rm Var}([p_{TA}]|N_A) &= \frac{|\Sigma|}{\Sigma_{11}}\langle [p_{TA}]\rangle^2. \label{varptA}
\end{align}
Both $\Sigma_{11}$ and $|\Sigma|$ are fitted to experimental data. 
%%This is a simplification. We don't fit the centrality dependence of sigma_11. 
On the other hand, the covariance $C_{12}$ is taken from the initial-state model, and we need to express $\Sigma_{12}$ as a function of $C_{12}$, $\Sigma_{11}$ and $|\Sigma|$. 
We first express $C$ as a function of $\Sigma$ using Eq.~(\ref{matrixform}):  
\begin{equation}
\label{covX2}
C=M^{-1} \Sigma M^{-1T}.
\end{equation}
The expression of $C_{12}$ yields a quadratic equation for $\Sigma_{12}$ of the form $a\Sigma_{12}^2+b \Sigma_{12}+c=0$, with 
\begin{align}
a&=\frac{D_A}{C_A} (1 + D_Ac_s^2)\frac{1}{\Sigma_{11}}\nonumber\\
b&=-(1+2D_Ac_s^2 )\\
c&= C_A c_s^2 \left(\Sigma_{11} -\frac{3}{2}C_{12}\right)+\frac{D_A}{C_A} (1 + D_Ac_s^2)\frac{|\Sigma|}{\Sigma_{11}}.\nonumber
\end{align}
The solution which does not diverge in the limit $D_A\to 0$ is
\begin{equation}
\Sigma_{12}=\frac{2c}{-b+\sqrt{b^2-4ac}}.
\end{equation}
We finally express the relative variance of $S$, $C_{11}$, and the relative variance of $R^2$,  $C_{22}$, as a function of $\Sigma_{11}$ and $|\Sigma|$ in the case where $C_{12}=0$: 
\begin{align}
C_{11}&=\frac{  \frac{1}{2}\Sigma_{11}+\sqrt{ \left(\frac{1}{2}\Sigma_{11}\right)^2-\frac{D_A^2}{C_A^2}(1+D_Ac_s^2)^2|\Sigma|  } }{(1+D_A c_s^2)^2}\nonumber    \\
\label{relvolume}
C_{22}&=   \left( \frac{2}{3C_Ac_s^2}\right)^2 \frac{|\Sigma|}{C_{11}}. 
\end{align}
If $D_A=0$, one recovers the simple result $C_{11}=\Sigma_{11}$: 
The relative entropy fluctuation equals the relative multiplicity fluctuation, according to Eq.~(\ref{smallfluct}). 
Eqs.~(\ref{relvolume}) allow us, by reverse engineering the hydrodynamic response, to infer the relative variances of $S$ and $R^2$ at $b=0$ from ATLAS data (Eqs.~(\ref{reverseengineering})). 

\section{Parametrization of the $b$ dependence}
\label{s:fittingfunctions}

Our model includes, in addition to the speed of sound $c_s$, five functions of $b$ which define the Gaussian distribution of $N_{A}$ and $p_{TA}$ at fixed $b$: 
\begin{itemize}
\item The mean and variance of $N_{ch}$, from which one infers those of $N_A$ as explained in Sec.~\ref{s:geometry}. 
The relative variance of $N_A$ is $\Sigma_{11}$ in the notations of Appendix~\ref{s:mappingdetails}. 
\item The mean  $\langle [p_{TA}]|b\rangle$. 
\item The variance of $[p_{TA}]$ at fixed $N_A$ (right-hand side of Eq.~(\ref{gaussianmoment2})). We parametrize the relative variance  $|\Sigma|/\Sigma_{11}$. 
\item The relative covariance of $S$ and $R^2$, $C_{12}$. 
\end{itemize} 
We specify the parametrizations we have chosen.
However, our results are robust with respect to this choice, which is to a large extent arbitrary. 

The mean multiplicity $\langle N_{ch}|b\rangle$ is parametrized as the exponential of a polynomial of degree 3 in the centrality fraction $c=\pi b^2/\sigma_{\rm PbPb}$~\cite{Das:2017ned}. 
For the mean transverse momentum, we choose a rational parametrization $\langle [p_{TA}]|b\rangle\propto (1+a_1c+a_2c^2)/(1+a_3 c^2)$. 
We have checked that a polynomial of degree 3 produces equivalent results. 

The remaining functions, $\Sigma_{11}$, $|\Sigma|/\Sigma_{11}$ and $C_{12}$, which are relative (co)variances, are all parametrized in the same way.  
We express them as a function of $x\equiv \langle N_{ch}|b\rangle/ \langle N_{ch}|0\rangle$, which is a ``multiplicity fraction'' relative to the knee. 
We write any of these three functions $f(x)$ in the form
\begin{equation}
\label{paramvariance}
f(x)=\frac{A_0+A_1(1-x)-A_2(1-x)^2}{x}, 
\end{equation}
where $A_0$, $A_1$ and $A_2$ are parameters specific to each function. 

The most important parameter for ultracentral collisions is the value at the knee, $f(1)=A_0$, and the other parameters $A_1$ and $A_2$ are of decreasing importance. 
The term proportional to $A_0$ is the relative variance of a Poisson distribution of mean proportional to $x$, while that proportional to $A_1$ describes the modification due to the binomial suppression of fluctuations~\cite{Bzdak:2012ab,Roubertie:2025qps}. 
The $A_2$ term is added to parametrize corrections to this simple picture. 

For $|\Sigma|/\Sigma_{11}$, all three parameters $A_0$, $A_1$ and $A_2$ are fitted to ATLAS data in Figs.~\ref{fig:ATLASfits} (b) and (c). 

For $\Sigma_{11}$, the fit to $P(N_{ch})$ gives $A_0\approx 1.15\times 10^{-3}$  after subtracting Poisson fluctuations (Sec.~\ref{s:geometry}), while $A_1$ and $A_2$ cannot be obtained from data. 
We assume that the centrality dependences of $\Sigma_{11}$ and $C_{11}$ are similar. 
For $C_{11}$, Eq.~(\ref{paramvariance}) with $A_2=0$ is equivalent to the parametrization proposed in Refs.~\cite{Samanta:2023kfk,Alqahtani:2024ejg}, which gives a good fit to initial-state models. 
We therefore set $A_2\equiv 0$, and the only remaining free parameter is $A_1/A_0$, which we vary between $1$ and $5$.  

For $C_{12}$, the default assumption is that it vanishes for $b=0$, that is, $A_0=0$, following the general arguments put forward in Ref.~\cite{Zhou:2025bwu}. 
If $A_0$ differs from $0$, this has a direct effect on $c_s^2$~\cite{Nijs:2023bzv}, which is quantified in Sec.~\ref{s:fits}. 
The parameters $A_1$ and $A_2$ are taken from an initial-state calculation described in Appendix~\ref{s:trento}. 
We take into account a possible variation of $C_{12}$ by $\pm 50\%$ in evaluating error bars. 

\begin{figure}[ht]
\includegraphics[width=\linewidth]{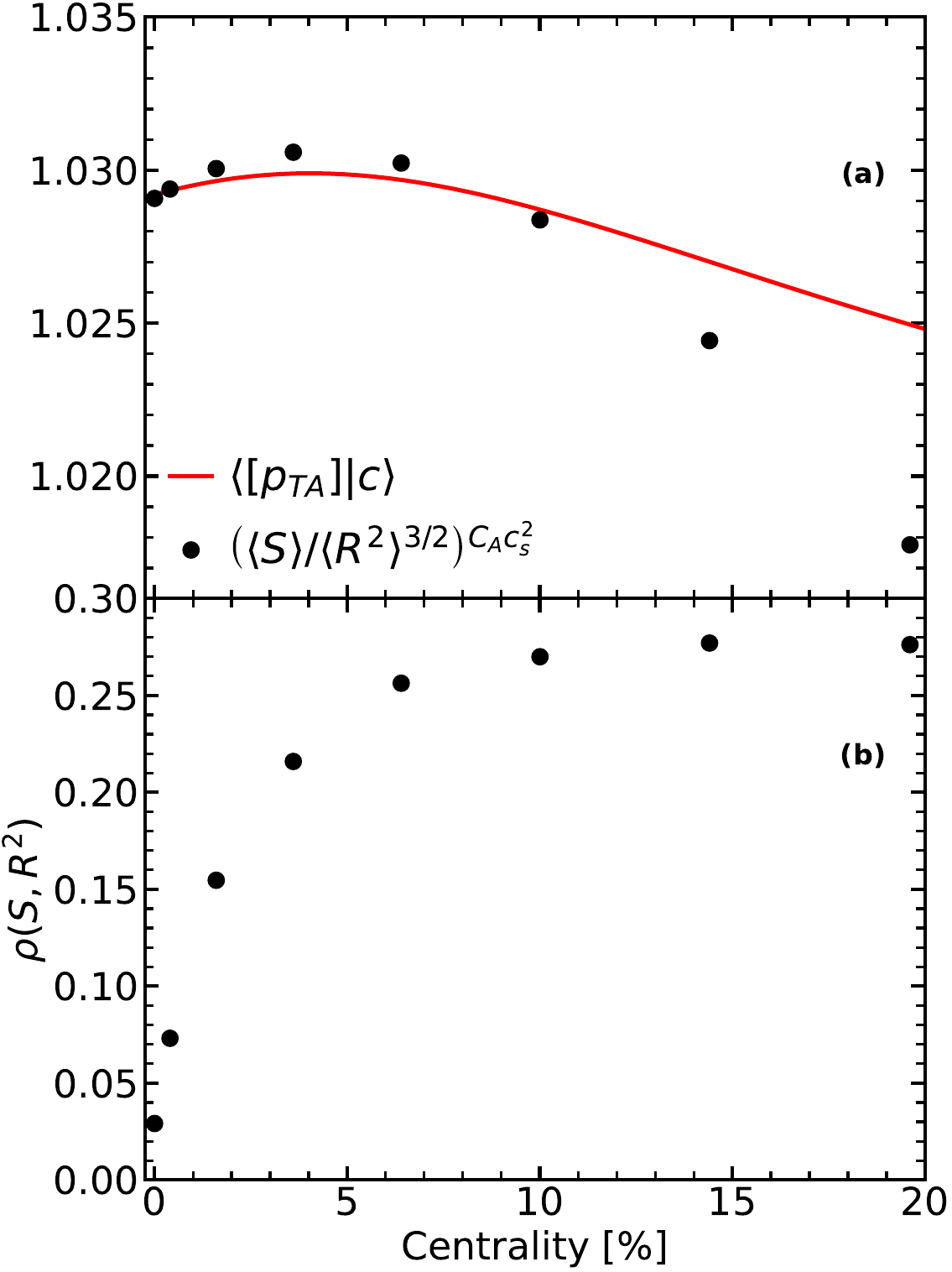}
    \caption{
    (a) Line: Variation of $\langle[p_{TA}]\rangle$, reconstructed from ATLAS data,  in GeV/$c$, as a function of the centrality fraction $c=\pi b^2/\sigma_{\rm PbPb}$. For the sake of illustration, we also display values in our initial-state simulation, assuming Eq.~(\ref{effhydro}), and taking into account the effect of the acceptance cuts. 
   (b) Pearson correlation coefficient between $S$ and $R^2$ at fixed $b$ in our initial-state simulation. 
}      
     \label{fig:trento}
\end{figure}

\section{Initial-state simulations}
\label{s:trento}
 
We present results of initial-state simulations using the popular \trento{} model~\cite{Moreland:2014oya}. 
The entropy density at a point ${\bf x}$  is evaluated as $s({\bf x})\propto \sqrt{T_A({\bf x})T_B({\bf x)}}$, where $T_{A,B}({\bf x})$ are the thickness functions of the incoming nuclei, evaluated by summing over all participant nucleons. 
This is the scenario favored by theory-to-data comparisons~\cite{Bernhard:2016tnd,Nijs:2020roc,JETSCAPE:2020mzn} and general theoretical arguments~\cite{Zhou:2025bwu}. 
We do not include nucleon substructure~\cite{Moreland:2018gsh} and choose a nucleon width of $0.5$~fm~\cite{Giacalone:2022hnz}. 
These details have little incidence on the quantities calculated in this Appendix. 
In the \trento{} model, event-by-event fluctuations come not only from the positions of nucleons, but also from fluctuations in the weight associated with each nucleon in evaluating the thickness functions, which are sampled according to a gamma distribution. 
We tune the parameter of this gamma distribution so as to match the relative entropy fluctuations at $b=0$ inferred from data, Eq.~(\ref{reverseengineering}). 
We find an inverse relative variance (of the gamma distribution) $k=4.5$, larger by a factor $\approx 2$ than the value favored by usual theory-to-data comparisons~\cite{Nijs:2020roc}. 
This is an interesting finding, which suggests that fluctuations from the collision process might be largely overestimated.   

We generate $5\times 10^5$ Pb+Pb collision events at fixed impact parameter $b$ for several values of $b$. 
The quantity of interest is the Pearson correlation $\rho$ between the total entropy $S$ and the mean square radius $R^2$ of the entropy density profile, which is displayed in Fig.~\ref{fig:trento} (b). 
It almost vanishes at $b=0$, which is a generic property of the $\sqrt{T_AT_B}$ ansatz~\cite{Zhou:2025bwu}. 
It then steeply increases with $b$, and a significant correlation is observed already at $5\%$ centrality. 
Our extraction of the speed of sound  relies on the hypothesis that $\rho$ vanishes at $b=0$, but we do take into account this steep increase. 
Specifically, we take the relative covariance between $S$ and $R^2$ from this initial-state calculation, which we shift in such a way that it vanishes exactly at $b=0$. 
We vary this covariance by $50\%$ in both directions, but it turns out to be a negligible source of uncertainty on $c_s^2$. 
Only the value at $b=0$ matters. 

We also display in Fig.~\ref{fig:trento} (a) the prediction of this initial-state calculation, supplemented with the effective hydrodynamic description (\ref{effhydro}), for the centrality dependence of the mean transverse momentum.
We choose the proportionality factor so as to match ATLAS data at $b=0$. 
We show this to illustrate that the initial-state calculation predicts a mild increase up to centrality $\approx 4\,\%$, which is also seen in data. 
%Our initial-state calculation overpredicts the centrality dependence of $\langle p_T\rangle$, 

\bibliography{ptfluct}
\end{document}